# Hydrogen Absorption at Room Temperature in Nanoscale Titanium Benzene Complexes

A.B. Phillips and B.S. Shivaram

Department of Physics
University of Virginia
Charlottesville, VA. 22901

Email: bss2d@virginia.edu

**ABSTRACT**

In this letter we report the first room temperature gravimetric measurements of hydrogen absorption in nanoscale titanium-benzene complexes formed by laser ablation in a benzene atmosphere in a UHV chamber. We are able to obtain a 6% by weight absorption as predicted by recent density functional theory based calculations under the conditions of low benzene pressure (35 milli-torr) and for sub-monolayer samples. For samples synthesized under higher benzene pressures we find a systematic degradation of the hydrogen absorption.

**1. Introduction:** The first demonstration that hydrogen can bind to transition metal atoms in molecular form was established by Kubas and co-workers[1]. To date more than 500 such "Kubas complexes" are known most of them binding just one hydrogen molecule with the exception of a few which can bind two di-hydrogen molecules[2]. This binding involves the forward charge donation from the s bond of $H_2$ to the partially filled d orbital of the transition metal atom and the back-donation from the metal atom to the s* antibonding orbital of $H_2$. More recently Zhao et al[3]. and Yildirim and Ciraci[4] have shown theoretically that it is possible for transition metal atoms supported on carbon containing precursors such as $C_5H_5$ (cyclopentadiene) and carbon nanotubes also to bind up to five $H_2$ molecules associatively in a similar manner. They also calculated an average binding energy of ~0.5 eV per $H_2$ molecule, which suggests desorption of the $H_2$ at a reasonably low temperature. This encouraging prediction has led other theoretical investigators to look for similar scenarios operating in other systems. In particular possibilities of storing hydrogen in record quantities in TM-ethylene complexes (14 weight %) and Li-$C_{60}$ complexes (13 weight t%) were discovered by Durgun et. al.[5] and Jena et. al[6]. respectively. In experiments in our laboratory with a sensitive picogravimeter we have been able to measure $H_2$ absorption in monolayer films of titanium-ethylene complexes lending strong support to such theoretical predictions[7]. In the case of a TM bound to a carbon complex the 18-electron rule has been invaluable in predicting the maximum number of hydrogen molecules that can be adsorbed[8]. Using these concepts Weck et al. have reported calculations on hydrogen absorption in complexes formed by transition metal atoms bound to different organic ring compounds[9]. Similar calculations have also been reported by Kiran et al.[10] in other carbon ring complexes such as $TiC_4H_4$ and $TiC_8H_8$. Very recently, Wadnerkar and Chaudhuri have also reported calculations for cyclopropene which is expected to show $H_2$ absorption at the 9% by weight level. In this letter we report the first experimental observation of hydrogen absorption in a titanium-benzene complex measured at room temperature.

There are many examples in the literature of the synthesis of transition metals complexed to carbon ring compounds. Silvon, Van Dam, and Skell,[11] used co-condensation of tungsten from a hot filament with benzene to prepare the bis-(benzene)tungsten complex, $(\eta-C_6H_6)_2W$. The same complex in larger quantities (0.5 gm/hour) could be obtained by co-condensation of tungsten vaporized by an electron beam in a benzene atmosphere of $10^{-5}$ torr at room temperature[12]. In such experiments the high thermal energies of the vaporized metal atoms is apparently not detrimental to formation of the complexes. Nb(benzene)$_2$ complexes have also been prepared through electron beam vaporization but with the samples being collected on a substrate held at 77 K[13]. Using solution chemistry methods it is possible to prepare a variety of TM-ring complexes including those that have multi-deckered rings[14]. However, to our knowledge no experimental work has been reported where hydrogen absorption studies on such complexes on the nanogram scale have been performed at room temperature. It is worth mentioning in this regard that there are many studies in the area of matrix isolation spectroscopy where several metal-benzene complexes have been observed[15,16]. However, such studies by their very nature require a cryogenic environment and preclude the gravimetric investigation of hydrogen absorption attributable exclusively to the metal-carbon ring complex[17].

**2. Experimental Details:** For the work reported here titanium-benzene complexes were synthesized in a ultra-high vacuum (UHV) chamber utilizing pulsed laser ablation techniques. Ultrapure benzene is employed as a precursor in the UHV chamber within which a rotating titanium target is ablated with the focused beam of a Nd:YAG laser operating at 10 Hz with a fluence that was varied between 15-65 mJ/sec/cm$^2$. The benzene vapor pressure prior to ablation was set in the chamber to vary from 15 millitorr to 65 millitorr. As the metal atoms emerge from the target after ablation they are slowed down by the benzene molecules to form complexes. At this juncture lacking the requisite spectroscopic tools in our apparatus we are unable to ascertain the precise nature of these complexes. However, in comparison with the previous studies mentioned above it is very likely we are forming Ti-$(C_6H_6)_2$. These complexes are collected on the face of a surface acoustic wave (SAW) quartz transducer. The transducer maintained at room temperature is employed both as a substrate to collect the complexes and also serves as a sensitive gravimeter. To achieve the level of sensitivity needed to detect hydrogen absorption in nanograms of the sample a high resolution spectrometer with digital feedback which can track the resonance frequency of the SAW sensor to better than a part per million is employed. With this method a change in mass less than 5 picograms can be measured[18]. This ability enables us to easily measure hydrogen uptake in a monolayer or less of a material or in aggregates of isolated nanoclusters scattered on the transducer. In the present letter hydrogen absorption studies on nanoscale titanium-benzene complexes synthesized as described above are reported.

**3. Results:** In figure 1 we show on the right the shift in the SAW resonator frequency as the Ti target is ablated (between the minutes of 2 and 12) and Ti-benzene complexes are collected. During this period the benzene precursor pressure is also monitored and this is shown on the left vertical axis. The drop in the benzene pressure implies the formation of a complex with titanium atoms. It is possible to perform a simple analysis to arrive at an estimate of the number of molecules of a given titanium-benzene complex we are collecting on the face of the transducer. Through previous experiments we have established a mass calibration of the 315 MHz SAW sensors employed in this work[19]. Using this calibration we ascertain that every 100 kHz of frequency shift corresponds to approximately 7.5 nanogram of material on the face of the transducer. Assuming that the benzene rings in the complex fall flat on the transducer and knowing the area of the transducer (3 mm2) we calculate a total of $1.2 \times 10^{14}$ molecules per monolayer. The mass of this monolayer for one titanium bound to each benzene molecule turns out to be 15.3 ng. Thus it is safe to state that we are depositing a monolayer or less in all the experiments reported here. Our samples are thus truly on the nanometer scale. If we consider other factors that might possibly alter the above argument we end up making the effective number of layers of the samples deposited even smaller. For example if we assume a sandwich complex, where a Ti atom is sandwiched between two benzene moieties we would arrive at a reduced coverage due to the increased "molecular weight" of the complex. In addition, the 3 mm$^2$ area mentioned above is simply the geometrical area of the transducer. However due to the roughness of the transducer surface this would translate to a larger effective area and hence reduce the coverage further. Although we are unable to establish precisely the nature of the complexes we are forming on the transducer face, we can make conclusions from previous experimental studies. It is known through prior laser ablation studies that there is significant metal specificity in the formation of multiple-decker benzene sandwich clusters.[20] Ablating vanadium is prone to yield multiple magic numbered species, $V_m(C_6H_6)_n$, with n-1 = m and n = 1,2,3, etc. However, in these studies both Ti and Cr were found to predominantly yield complexes with n = 2.

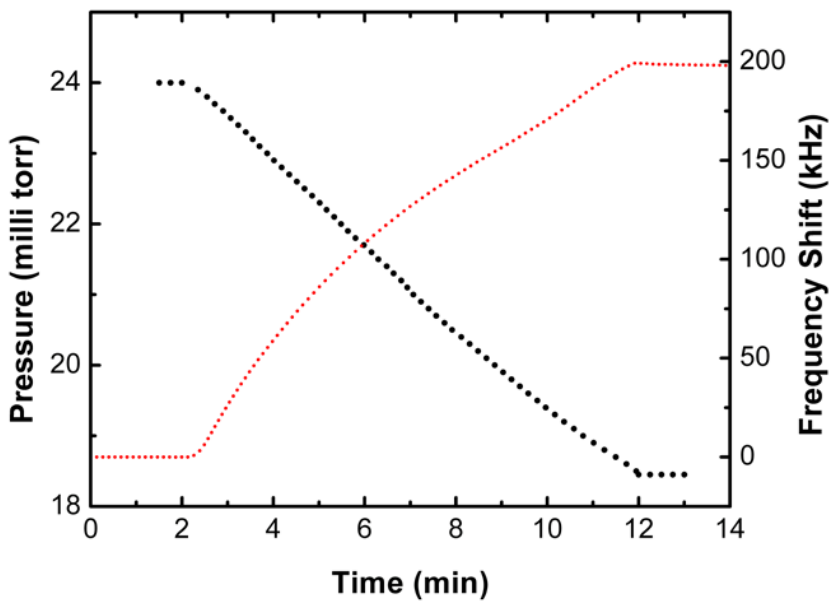

*Figure 1: Shows the frequency shift (red line) of the surface acoustic wave resonator as the Ti-benzene complex formed through laser ablation accumulates on its face. This frequency shift corresponds to 6000 pulses. Using the calibration of our sensor from ref.(xx) the total accumulated mass on the sensor is xx nanogram. The right vertical axis shows the drop in the benzene pressure as the titanium is ablated. In all the data reported in this letter we used half to quarter the number of pulses and hence the observed frequency shifts and the drop in benzene pressure are correspondingly lower.*

In figure 2 we show the wt % uptake of $H_2$ in a Ti-benzene complex synthesized at three different benzene pressures. In each experimental run we create and measure two separate samples and both the samples yield identical results. The results shown in the figure are from three different runs. We note that the $H_2$ uptake is large and appears to saturate at the 6 wt.% level. To obtain these results we had to trim down the number of laser pulses so that the samples created correspond to less than 5 nanograms. The hydrogen gas loading pressure in these experiments reaches a final value of 800 torr and the loading time is about 10 minutes starting from zero pressure. Also notable in figure 2 is the rapid kinetics of the hydrogen absorption.

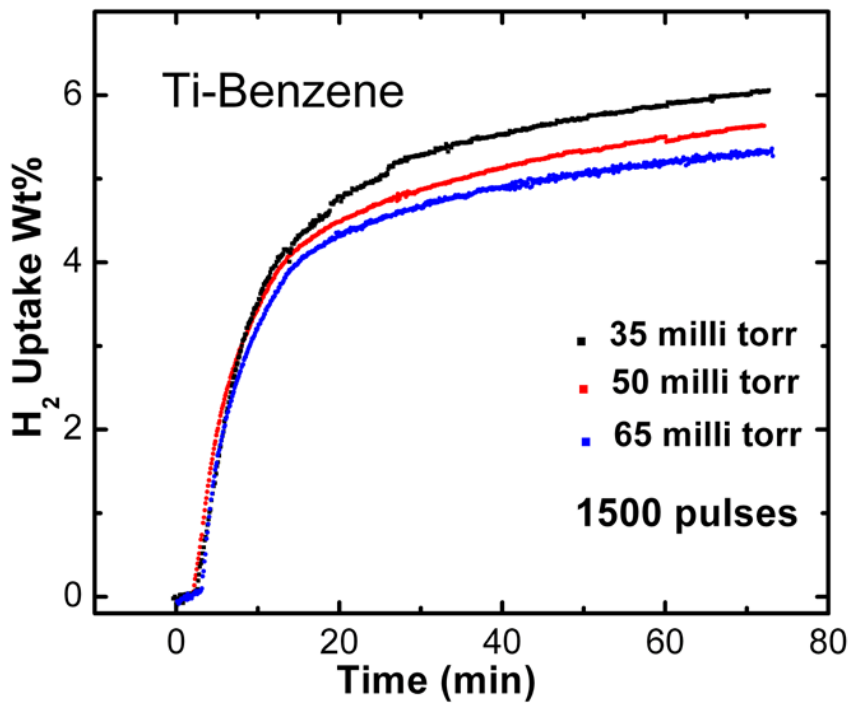

*Figure 2: Shows the wt % uptake at room temperature of $H_2$ in Ti-benzene complexes synthesized by ablating titanium in a benzene atmosphere of pressure 35, 50 and 65 millitorr. A Nd:YAG laser with a fluence of xxxx J/cm2 was used to ablate the metal and the complexes were collected in a period of 2-3 min (corresponding to 1500 laser pulses at 10 Hz).*

However, when we collect samples by ablating in a higher benzene precursor pressure we obtain a systematically reduced hydrogen absorption. In figure 3 we show the uptake obtained in various samples synthesized under benzene pressures ranging from 15 mtorr to 65 m torr. These results are similar to those we have reported earlier with the Ti-ethylene complexes where a degradation in hydrogen uptake was observed for samples synthesized at higher ethylene pressures[21]. It is possible that higher benzene pressures result in the formation of alternate nanostructures, perhaps nanoparticles, wherein the titanium atoms have a chance to aggregate. Such clustering effects in transition metals have been shown through many theoretical studies in the context of TM atoms complexed to buckyballs[22] and carbon nanotubes[23] to result in reduced $H_2$ uptake.

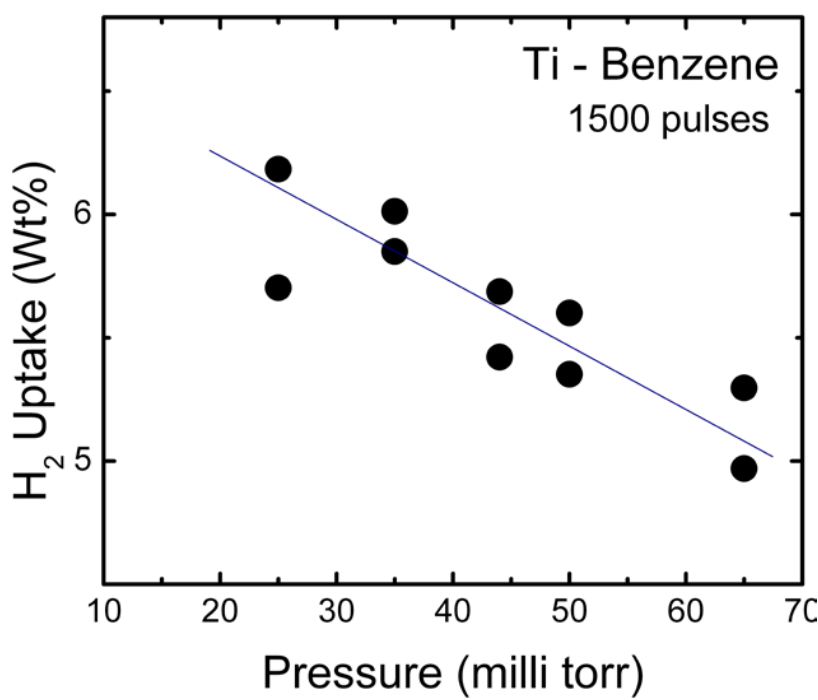

*Figure 3: Shows the dependence of the total hydrogen uptake for Ti-benzene complexes synthesized at various benzene precursor pressures. A qualitatively similar dependence is observed in Ti-ethylene complexes (ref.21). The line shown is a guide to the eye.*

**Concluding Remarks:** In work reported here we have observed molecular hydrogen absorption upto 6 weight percent at room temperature in Ti-benzene complexes synthesized in a pulsed laser ablation apparatus. This absorption furthermore occurs with rapid kinetics. These observations corroborate suggestions from recent calculations that titanium-benzene and other ring compounds forming complexes with metal atoms could exhibit hydrogen uptake upto 9% by weight at room temperature. Spectroscopic and high resolution electron microscopy studies to pin down the molecular and nanoscale structure of the titanium-benzene complexes studied in this work are currently being planned.

**Acknowledgements:** This work was supported partially from funds from NSF DMR 0838016. One of us (ABP) acknowledges support from a DOE fellowship administered by SURA.